\documentclass[twocolumn,aps,superscriptaddress,showpacs,nofootinbib,floatfix]{revtex4-1}   	% use "amsart" instead of "article" for AMSLaTeX format
\usepackage{graphicx}				% Use pdf, png, jpg, or eps§ with pdflatex; use eps in DVI mode
\usepackage{amsmath}								% TeX will automatically convert eps --> pdf in pdflatex		
\usepackage{amssymb}
\usepackage{soul}
\usepackage{enumitem}

\usepackage[normalem]{ulem}
\usepackage{multirow}
\usepackage{xcolor}

\renewcommand\sout{\bgroup \color{red} \ULdepth=-.5ex \ULset}

\begin{document}

%%%%%%%%%%%%%%%%%%% Title %%%%%%%%%%%%%%%%

\title{The structure of the $X(3915)$ meson and its production in heavy ion collisions}
			
\author{Sungtae Cho}
\email{sungtae.cho@kangwon.ac.kr}
\affiliation{Division of Science Education, Kangwon National University, Chuncheon 24341, Korea} 

\author{Aaron Park}
\email{aaron.park@yonsei.ac.kr}
\affiliation{Department of Physics and Institute of Physics and Applied Physics, Yonsei University, Seoul 03722, Korea}

\author{Su Houng Lee}%
\email{suhoung@yonsei.ac.kr}
\affiliation{Department of Physics and Institute of Physics and Applied Physics, Yonsei University, Seoul 03722, Korea}

\author{Sungsik Noh}
\email{sungsiknoh@kangwon.ac.kr}
\affiliation{Division of Science Education, Kangwon National University, Chuncheon 24341, Korea}

\begin{abstract}
We study the structure of the $X(3915)$ meson in a quark model and explore how its  production in heavy ion collisions depends on its internal structure.
We first analyze the $X(3915)$ as a $c\bar{c}s\bar{s}$ state and solve the Hamiltonian with color-spin interactions within the quark model.
We find that the ground state of the $c\bar{c}s\bar{s}$ with total spin 0 obtained from the quark model analysis favors a separated $D_s \bar{D}_s$ state.
To probe its structure further, we study its production in relativistic heavy ion collisions for various proposed configurations.
We calculate the transverse momentum distributions and yields for the $X(3915)$ assuming its structure to be either a charmonium, a tetraquark, or a hadronic molecular state.
We argue that by measuring the transverse momentum distributions and yields of the $X(3915)$ produced in heavy ion collisions, one can identify the structure of the $X(3915)$.
\end{abstract}

\maketitle

\section{Introduction}

Since the $X(3872)$ meson was observed in 2003 by the Belle Collaboration \cite{Belle:2003nnu}, many new hadrons with the names $X$, $Y$ and $Z$ have been observed \cite{Esposito:2016noz,Brambilla:2019esw,Chen:2022asf}.
The $X(3915)$ meson discovered by the Belle Collaboration in 2004 \cite{Belle:2004lle} from the analysis of
the $\omega J/\psi$ invariant mass in the $B$ meson decay is one of those $XYZ$ hadrons.
The $X(3915)$ has also been observed by the BaBar Collaboration \cite{BaBar:2007vxr,BaBar:2010wfc}, and its
quantum number could be determined as $I^G(J^{PC})=0^+(0^{++})$ by the Belle Collaboration and subsequently confirmed by the BaBar Collaboration \cite{Belle:2009and,BaBar:2012nxg}.
As a result, it is listed and renamed in the Particle Data Group as the $\chi_{c0}(3915)$ \cite{ParticleDataGroup:2024cfk}.

The mass of the $X(3915)$ meson is close to that of the $X(3872)$, but the two states differ significantly in their properties.
The $X(3915)$ has quantum numbers $0^+(0^{++})$, in contrast to the $X(3872)$, which has $0^+(1^{++})$.
Unlike the $X(3872)$, which decays dominantly into $\bar{D}^{*0}D^0$, the $X(3915)$ lies below the $D_s\bar{D}_s$ threshold but is considered capable of decaying into these mesons, analogous to the $X(3960)$ recently observed by LHCb Collaboration \cite{LHCb:2022aki}, and possibly having strange quarks as its constituents.

Like most $XYZ$ hadrons including the $X(3872)$ meson, the
internal structure of the $X(3915)$ meson has not been clearly
understood to date, and the proposed structures of the $X(3915)$
include an $s$-wave molecular bound state of the $D_s\bar{D}_s$
pair \cite{Li:2015iga}, a compact
$c\bar{c}s\bar{s}$ tetraquark state \cite{Lebed:2016yvr}, a conventional $c\bar{c}$ state
\cite{Duan:2020tsx}, and a hadronic molecular state with the
probability of a bare $c\bar{c}$ state less than 45\%
\cite{Ortega:2017qmg}.

Regarding the $X(3872)$, there have been attempts to understand the structure of the $X(3872)$ \cite{Suzuki:2005ha,Dong:2009yp,Takizawa:2012hy}, including studies in the quark model \cite{Yun:2022evm} and in heavy ion collisions \cite{Cho:2010db, Cho:2011ew, Cho:2017dcy}. 
Moreover, there have also been general studies on the possibilities of the bound state of exotic hadrons composed of four quarks, in various combinations in the quark model with color-spin interactions, focusing on doubly heavy exotic hadrons such as $T_{cc}$ \cite{Park:2018wjk, Noh:2021lqs, Noh:2023zoq, Park:2024gbq}.
These studies have significantly advanced our understanding of multiquark states in open heavy flavor sectors.
However, despite these extensive efforts, the internal structure of hidden charm and hidden strangeness tetraquarks with $c\bar{c}s\bar{s}$ configuration remains much less well understood than that of open heavy flavor tetraquarks.
In particular, the calculation in Ref.~\cite{Wu:2016gas} was limited to the color-spin interaction only. The authors of Ref.~\cite{Liu:2024fnh} employed a full model Hamiltonian, however, their predicted mass for the $c\bar{c}s\bar{s}$ configuration does not reproduce the experimentally measured mass of the $X(3915)$.

In parallel, detailed analyses of doubly heavy tetraquark states such as $T_{cc}$ and $T_{bb}$, which contain open heavy flavors, have shown that their internal structures are typically dominated by color triplet components due to strong correlation of the light diquark, with distinct features arising from different admixtures of color triplet and color sextet configurations \cite{Noh:2023zoq}. 
These results highlight the crucial role played by color-spin interactions in binding quarks together in tetraquark configurations.
However, the $X(3915)$ differs fundamentally from these states in that it possesses a hidden flavor $c\bar{c}s\bar{s}$ quark composition, for which diquark correlations are expected to be suppressed due to the masses of heavy quarks.
Although its mass lies close to a relevant hadronic decay threshold, similar to the case of the $T_{cc}$, the absence of open heavy flavors and the presence of both charm and strange quarks may lead to qualitatively different color-spin dynamics and internal color configurations.
As a consequence, the internal structure of the $X(3915)$ cannot be directly inferred from the patterns observed in doubly heavy open flavor tetraquarks.
A dedicated investigation of the color-spin structure specific to a $c\bar{c}s\bar{s}$ configuration associated with the $X(3915)$, together with an investigation of how different internal structures give rise to distinct experimentally accessible observables, is therefore essential.

In this work, we employ a quark model based on a full model Hamiltonian that includes a Yukawa type hyperfine potential, which has been shown to reproduce hadron masses accurately even when a single dominant spatial basis is adopted \cite{Park:2024gbq}.
We calculate the mass of the $X(3915)$ and precisely reproduce the experimentally measured value, thereby providing a more reliable foundation for investigating its internal structure.

Furthermore, we investigate the internal structure and production
of the $X(3915)$ in relativistic heavy ion collisions based on a
coalescence model, by calculating the transverse momentum
distributions and yields for various proposed states of the
$X(3915)$, i.e, a charmonium, a tetraquark, and a hadronic
molecular state. As it has been known that the yield of exotic
hadrons strongly depends on their internal structures
\cite{Cho:2010db, Cho:2011ew, Cho:2017dcy}, we expect to be able
to provide a detailed assessment of whether the $X(3915)$ behaves
more like a hadronic molecular or a compact multiquark state by
analyzing the internal color-spin structure and studying
production in relativistic heavy ion collisions through a
coalescence model. Compared to previous studies, our approach
explicitly connects the internal quark configuration to observable
production characteristics, offering a unique perspective on the
nature of the $X(3915)$.

Here, we assume that the $X(3915)$ exists in a $2p$ state when the
$X(3915)$ meson is produced as a charmonium state, though its
internal structure is still unclear, and its existence as a $2p$
state, the $\chi_{c0}(2P)$ has been refuted because of inconsistencies in its decay modes and 
widths compared to theoretical evaluations \cite{Guo:2012tv,
Olsen:2014maa}. Concerning the
production of the $X(3915)$ in a tetraquark state, we regard the
$X(3915)$ as the $c\bar{c}s\bar{s}$ state, while we consider the
$D_s\bar{D}_s$ for the $X(3915)$ in a hadronic molecular state.

The paper is organized as follows. In Sec.~II, we first briefly
describe the simplified quark model. In Sec.~III, we then analyze
internal color-spin structure of the $X(3915)$ based on the quark
model calculations. In Sec.~IV, we introduce the yield and
transverse momentum distribution within the coalescence model to
consider the production of $X(3915)$ mesons by recombination
in relativistic heavy ion collisions
for three possible states of the $X(3915)$, i.e., a charmonium, a
tetraquark, and a hadronic molecular state. Then, in Sec.~V we
calculate the yield and transverse momentum distribution of the
$X(3915)$, and present the transverse momentum distribution ratio
between the $X(3915)$ and $D_s$. We also discuss the possibilities
of identifying the internal structure of the $X(3915)$ meson from
its production in heavy ion collisions in Sec.~V. The Sec.~VI is
devoted to conclusions.

\section{Quark model description}
To analyze the short range part of the interaction, we adopt a simplified quark model described in Ref.~\cite{Park:2018wjk}.
However, as shown in Refs.~\cite{Noh:2023zoq,Park:2024gbq}, the Yukawa type hyperfine potential is more suitable for describing short range interactions than the Gaussian type used in Ref.~\cite{Park:2018wjk}.
In particular, as presented in Ref.~\cite{Park:2024gbq}, the Yukawa type potential provides a good fit to hadron masses even when using only the most dominant single spatial basis in the calculations.
We thus calculate the tetraquark state using only the most dominant single spatial basis and adopt the Yukawa type hyperfine potential from Ref.~\cite{Park:2024gbq}, instead of the Gaussian type used in Ref.~\cite{Park:2018wjk}.
With the Yukawa type hyperfine potential, the simplified quark model is described by the following Hamiltonian.
\begin{align}
H = \sum^{4}_{i=1} \left( m_i+\frac{{\vec p}^{\,\,2}_i}{2 m_i} \right)-\frac{3}{4}\sum^{4}_{i<j}\frac{\lambda^{c}_{i}}{2} \,\, \frac{\lambda^{c}_{j}}{2} \left( V^{C}_{ij} + V^{CS}_{ij} - D \right),
\label{Hamiltonian}
\end{align}
where the confinement and the Yukawa type hyperfine potentials are given as follows:
\begin{align}
V^{C}_{ij} &= - \frac{\kappa}{r_{ij}} + \sigma \, r_{ij},
\label{ConfineP}
\\
V^{CS}_{ij} &= \frac{\mu_{ij} \mu'_{ij}}{m_i m_j c^4} \frac{\hbar \, c}{r_{ij}} e^{- \mu_{ij} r_{ij}/\hbar c} \, \vec{\sigma}_i \cdot \vec{\sigma}_j.
\label{CSP}
\end{align}
Here, $m_i$ represents the quark mass, $\lambda^c_{i}/2$ and $\vec{\sigma}_i$ denote the SU(3) color and SU(2) spin operators, respectively, for the $i$th quark.
$r_{ij}\equiv|{\vec r}_i - {\vec r}_j |$ denotes the relative distance between the $i$th and $j$th quarks. Also, $\mu_{ij}$ and $\mu'_{ij}$ contain additional mass dependences as follows:
\begin{align}
\mu_{ij} &= \left( \alpha + \beta \frac{m_i m_j}{m_i + m_j} \right)	,	
\label{Parameter1}
\\
\mu'_{ij} &= \left( \gamma+ \delta \frac{m_i m_j}{m_i + m_j} \right)	.	
\label{Parameter2}
\end{align}
We adopt the model fitting parameters determined from calculations using only the most dominant single spatial basis, as presented in Ref.~\cite{Park:2024gbq}, as follows:
\begin{align}
&\kappa=107.7 \, \textrm{MeV fm}, \,\,  \sigma=933.271 \, \textrm{MeV/fm}, \,\, D=950  \, \textrm{MeV},&
\nonumber \\
&m_{u}=320 \, \textrm{MeV}, \quad m_{s}=612 \, \textrm{MeV}, \quad m_{c}=1893 \, \textrm{MeV}, &
\nonumber \\
&\alpha = 223.946 \, \textrm{MeV}, \,\, \beta = 0.227992,& \nonumber \\
& \gamma=230.244 \, \textrm{MeV}, \,\, \,\, \delta = 0.315434,	 &
\label{SimpleFitParameters}
\end{align}
where we re-define a few parameters, compared to those of Ref.~\cite{Park:2024gbq}.
A few of the fitting results using the parameter set in Eq.~(\ref{SimpleFitParameters}) are shown in Table~\ref{FitResult}. The full results for fitting to hadrons are found in Tables IV and V of Ref.~\cite{Park:2024gbq}.
%%%%%%%%%%%%%%%%%%%%%%%%%%%		Table 1. Meson and Baryon masses		%%%%%%%%%%%%%%%%%%%%%%%%%%%%%%%%%%%%%%
\begin{table}[t]

\caption{This table shows the fitting results for a few of the hadron masses that are taken directly from Ref.~\cite{Park:2024gbq} and obtained(Column 3) using the most dominant single spatial basis and the parameter set given in Eq.~(\ref{SimpleFitParameters}). Column 4 shows the variational parameters $a_1$ and $a_2$. The results for all the other mesons and baryons are referred to Tables IV and V of Ref.~\cite{Park:2024gbq}.}	

\centering

\begin{tabular}{ccccc}
\hline
\hline	\multirow{2}{*}{Particle}	&	Experimental	&	Mass		&	Variational			& Error			\\
								&	Value (MeV)	&	(MeV)	&	Parameters (${\rm fm}^{-2}$)	&(\%)	\\
\hline 
$D$			&	1864.8		&	1860.4		&	$a_1$ = 4.4	&0.24	\\
$D^*$		&	2007.0		&	2005.2  		&	$a_1$ = 3.5	&0.08	\\
$\eta_{c}$	&	2983.6		&	2995.4		&	$a_1$ = 14.3	&0.39	\\
$J/\Psi$		&	3096.9		&	3117.5		&	$a_1$ = 11.0	&0.67	\\
$D_s$		&	1968.3		&	1961.0		&	$a_1$ = 7.0	&0.37	\\
$D^*_s$		&	2112.1		&	2097.6		&	$a_1$ = 5.4	&0.69	\\

$\Lambda_{c}$	&	2286.5	&	2265.4	&	\quad$a_1$ = 2.7, $a_2$ = 3.5\quad	&0.92	\\
$\Xi_{cc}$		&	3621.4	&	3609.6	&	\quad$a_1$ = 7.3, $a_2$ = 3.0\quad	&0.33	\\
$\Sigma_{c}$		&	2452.9	&	2444.6	&	\quad$a_1$ = 2.0, $a_2$ = 3.5\quad	&0.37	\\
$\Sigma_{c}^*$	&	2517.5	&	2528.2	&	\quad$a_1$ = 1.8, $a_2$ = 3.1\quad	&0.39	\\
$\Xi_{c}$		&	2467.8	&	2459.5	&	\quad$a_1$ = 3.1, $a_2$ = 4.4\quad	&0.44	\\
$\Xi_{c}^*$		&	2645.9	&	2639.9	&	\quad$a_1$ = 2.3, $a_2$ = 4.0\quad	&0.24	\\
\hline 
\hline
\label{FitResult}
\end{tabular}
\end{table}
%%%%%%%%%%%%%%%%%%%%%%%%%%%%%%%%%%%%%%%%%%%%%%%%%%%%%%%%%%%%%%%%%%%%%%%%%%%%%%%%%%%%%%%%

The Jacobi coordinates describing the four-quark configuration, labelled by $\bar{c}(1)\bar{s}(2) c(3) s(4)$, can be set as follows:
\begin{enumerate}[label=(\alph*)]
	\item{Coordinates Set 1}
	\begin{align}
	& \vec{x}_1 = \frac{1}{\sqrt{2}}({\vec r}_1 - {\vec r}_4), \qquad \vec{x}_2 = \frac{1}{\sqrt{2}}({\vec r}_2 - {\vec r}_3)\,, 	\nonumber
	\\
	& \vec{x}_3 = \frac{1}{M_\mu} \left( \frac{m_1 {\vec r}_1 + m_4 {\vec r}_4}{m_1 + m_4} - \frac{m_2 {\vec r}_2 + m_3 {\vec r}_3}{m_2 + m_3} \right)\,,
\label{CoordSet1}
	\end{align}
	\item{Coordinates Set 2}
	\begin{align}
	& \vec{y}_1 = \frac{1}{\sqrt{2}}({\vec r}_1 - {\vec r}_3), \qquad \vec{y}_2 = \frac{1}{\sqrt{2}}({\vec r}_4 - {\vec r}_2)\,, 	\nonumber
	\\
	& \vec{y}_3 = \frac{1}{M_\mu} \left( \frac{m_1 {\vec r}_1 + m_3 {\vec r}_3}{m_1 + m_3} - \frac{m_2 {\vec r}_2 + m_4 {\vec r}_4}{m_2 + m_4} \right)\,, 
\label{CoordSet2}
	\end{align}
	\item{Coordinates Set 3}
	\begin{align}
	& \vec{z}_1 = \frac{1}{\sqrt{2}}({\vec r}_1 - {\vec r}_2), \qquad \vec{z}_2 = \frac{1}{\sqrt{2}}({\vec r}_3 - {\vec r}_4)\,, 	\nonumber
	\\
	& \vec{z}_3 = \frac{1}{M_\mu} \left( \frac{m_1 {\vec r}_1 + m_2 {\vec r}_2}{m_1 + m_2} - \frac{m_3 {\vec r}_3 + m_4 {\vec r}_4}{m_3 + m_4} \right)\,, 
\label{CoordSet3}
	\end{align}
\end{enumerate}
where
\begin{align}
M_\mu = \left[ \frac{m_1^2 + m_4^2}{(m_1 + m_4)^2} + \frac{m_2^2 + m_3^2}{(m_2 + m_3)^2} \right]^{1/2} \,,	\nonumber
\end{align}
and
\begin{align}
&m_1=m_3=m_c,	\,\,	m_2=m_4=m_s& 	\quad {\rm for} \,\, c\bar{c}s\bar{s}. \nonumber
\end{align}
We choose coordinates set 1 as our reference, as it is suitable for studying open flavor decay mode $D_s \bar{D}_s$ of the $X(3915)$.
Using the transformations between the different sets of Jacobi coordinates introduced above, we calculate the relevant matrix elements involving two quarks.
We construct the spatial wave function in a simple Gaussian form, as follows:
\begin{align}
&\psi^{Spatial}(\vec{x}_1, \vec{x}_2, \vec{x}_3)
\nonumber \\
&=
\left( \frac{2}{\pi} \right)^{\frac{9}{4}} a_1^{\frac{3}{4}} a_1^{\frac{3}{4}} a_3^{\frac{3}{4}} \exp \big[- a_1 \vec{x}_1^{\,2} - a_1 \vec{x}_2^{\,2} - a_3 \vec{x}_3^{\,2} \big],
\label{spatial}
\end{align}
where the variational parameter is set as $a_2=a_1$ to ensure that the spatial wave function is symmetric under the permutation $(13)(24)$.

The color-spin(CS) wave function for the $X(3915)$, which has quantum numbers $I^G(J^{PC})=0^+(0^{++})$, is composed of four CS basis states: $\left|CS'_1\right\rangle=\left|(\bar{c}c)^{\mathbf{1}}_0 (\bar{s}s)^{\mathbf{1}}_0 \right\rangle, \left|CS'_2\right\rangle=\left|(\bar{c}c)^{\mathbf{8}}_0 (\bar{s}s)^{\mathbf{8}}_0 \right\rangle, \left|CS'_3\right\rangle=\left|(\bar{c}c)^{\mathbf{1}}_1 (\bar{s}s)^{\mathbf{1}}_1 \right\rangle, \left|CS'_4\right\rangle=\left|(\bar{c}c)^{\mathbf{8}}_1 (\bar{s}s)^{\mathbf{8}}_1 \right\rangle$, where $c$ and $s$ denote charm and strange quarks, respectively.
The superscripts(subscripts) indicate the color(spin) states of each quark-antiquark pair.
These CS bases can be represented through a different quark-antiquark basis, $(\bar{c}s) \otimes (\bar{s}c)$, as follows:
\begin{align}
\left|CS'_1\right\rangle
&=
-\frac{1}{2\sqrt{3}} \left|CS_1\right\rangle + \frac{1}{6} \left|CS_2\right\rangle - \sqrt{\frac{2}{3}} \left|CS_3\right\rangle + \frac{\sqrt{2}}{3} \left|CS_4\right\rangle,
\nonumber \\
\left|CS'_2\right\rangle
&=
-\sqrt{\frac{2}{3}} \left|CS_1\right\rangle + \frac{\sqrt{2}}{3} \left|CS_2\right\rangle + \frac{1}{2 \sqrt{3}} \left|CS_3\right\rangle - \frac{1}{6} \left|CS_4\right\rangle,
\nonumber \\
\left|CS'_3\right\rangle
&=
-\frac{1}{6} \left|CS_1\right\rangle - \frac{1}{2 \sqrt{3}} \left|CS_2\right\rangle - \frac{\sqrt{2}}{3} \left|CS_3\right\rangle - \sqrt{\frac{2}{3}} \left|CS_4\right\rangle,
\nonumber \\
\left|CS'_4\right\rangle
&=
-\frac{\sqrt{2}}{3} \left|CS_1\right\rangle - \sqrt{\frac{2}{3}} \left|CS_2\right\rangle + \frac{1}{6} \left|CS_3\right\rangle + \frac{1}{2 \sqrt{3}} \left|CS_4\right\rangle,
\label{CSbases}
\end{align}
where $\left|CS_1\right\rangle
=
\left|(\bar{c}s)^{\mathbf{1}}_1 (\bar{s}c)^{\mathbf{1}}_1 \right\rangle$, $\left|CS_2\right\rangle
=
\left|(\bar{c}s)^{\mathbf{1}}_0 (\bar{s}c)^{\mathbf{1}}_0 \right\rangle$, $\left|CS_3\right\rangle
=
\left|(\bar{c}s)^{\mathbf{8}}_1 (\bar{s}c)^{\mathbf{8}}_1 \right\rangle$, and $\left|CS_4\right\rangle
=
\left|(\bar{c}s)^{\mathbf{8}}_0 (\bar{s}c)^{\mathbf{8}}_0 \right\rangle$.
However, for the analysis of the open flavor decay channel $D_s \bar{D}_s$, it is more appropriate to represent the four CS states in the $(\bar{c}s) \otimes (\bar{s}c)$ basis, which better reflects the structure of the final state of the $D_s \bar{D}_s$ mesons.
This can be obtained through an orthogonal transformation from the original basis set of Eq.~(\ref{CSbases}).
The transformed CS basis states are given as follows:
\begin{align}
\left|CS_1\right\rangle
&=
-\frac{1}{2\sqrt{3}} \left|CS'_1\right\rangle - \sqrt{\frac{2}{3}} \left|CS'_2\right\rangle - \frac{1}{6} \left|CS'_3\right\rangle - \frac{\sqrt{2}}{3} \left|CS'_4\right\rangle,
\nonumber \\
\left|CS_2\right\rangle
&=
\frac{1}{6} \left|CS'_1\right\rangle + \frac{\sqrt{2}}{3} \left|CS'_2\right\rangle - \frac{1}{2 \sqrt{3}} \left|CS'_3\right\rangle - \sqrt{\frac{2}{3}} \left|CS'_4\right\rangle,
\nonumber \\
\left|CS_3\right\rangle
&=
-\sqrt{\frac{2}{3}} \left|CS'_1\right\rangle + \frac{1}{2 \sqrt{3}} \left|CS'_2\right\rangle - \frac{\sqrt{2}}{3} \left|CS'_3\right\rangle + \frac{1}{6} \left|CS'_4\right\rangle,
\nonumber \\
\left|CS_4\right\rangle
&=
\frac{\sqrt{2}}{3} \left|CS'_1\right\rangle - \frac{1}{6} \left|CS'_2\right\rangle - \sqrt{\frac{2}{3}} \left|CS'_3\right\rangle + \frac{1}{2 \sqrt{3}} \left|CS'_4\right\rangle.
\label{CSbases2}
\end{align}

\section{Short range interaction}

As shown in Table~\ref{TetraResult}, the quark model calculation for the Hamiltonian in Eq.~(\ref{Hamiltonian}) gives the ground state mass of 3922.1 MeV, which exactly reproduces the measured mass of the $X(3915)$.
The parameters $a_1$ and $a_3$ are determined through a variational method applied to the ground state.
As the variational parameter $a_3$ goes to $0$, the ground state mass of the $X(3915)$ approaches its threshold value.
Furthermore, the parameter $a_1$ is the same as that for the $D_s$ meson given in Table~\ref{FitResult}.
This implies that the tetraquark wave function reduces to those of its threshold.
\begin{widetext}

%%%%%%%%%%%%%%%%%%%%%%%%%%%		Table 2. Tetra Result		%%%%%%%%%%%%%%%%%%%%%%%%%%%%%%%%%%%%%%
\begin{table}[ht]

\caption{Masses of the $X(3915)$ and the corresponding thresholds, calculated using the full model Hamiltonian given in Eq.~(\ref{Hamiltonian}). The threshold masses are obtained from the masses of the threshold mesons presented in Table~\ref{FitResult}.}

\centering

\begin{tabular}{cccccc}
\hline
\hline
Configuration		&	$J^{PC}$	&Threshold(MeV)& Measured mass(MeV)	&	Mass(MeV)	&	Variational parameters (${\rm fm}^{-2}$)	\\
\hline 
$c\bar{c}s\bar{s}$	&	$0^{++}$	& $D_s \bar{D}_s$(3922.0)	&	3922.1		&	3922.1		&	$a_1$ = 7.0, $a_3$ = 0.001	\\
\hline 
\hline
\label{TetraResult}
\end{tabular}
\end{table}
%%%%%%%%%%%%%%%%%%%%%%%%%%%%%%%%%%%%%%%%%%%%%%%%%%%%%%%%%%%%%%%%%%%%%%%%%%%%%%%%%%%%%%%%
\end{widetext}

To investigate the short distance part of interaction, we adopt the color-spin factor $K$, defined in Ref.~\cite{Yun:2022evm}, as follows:
\begin{align}
  K&=-\sum_{i<j}^n \frac{1}{m_im_j} \lambda^c_i \lambda^c_j  \vec{\sigma}_i \cdot \vec{\sigma}_j,
\label{color-spin0} 
\end{align}
where $\lambda_i^c$ and $\sigma_i$ are the color and spin operators, respectively, and $m_i$ the mass of the $i$th quark among the $n$ quarks.
The $K$ factor for the $X(3915)$ in the CS bases given in Eq.~(\ref{CSbases2}) is as follows:
\begin{widetext}
$K_{X(3915)}=$
\begin{align}
\small
\left(
\begin{array}{cccc}
\frac{16}{3} \left( \frac{1}{m_{\bar{c}} m_s} + \frac{1}{ m_{\bar{s}} m_c} \right) & 0 & \frac{8 \sqrt{2} \left( m_{\bar{c}} - m_s \right) \left( m_{\bar{s}} - m_c \right)}{3 m_{\bar{c}} m_{\bar{s}} m_c m_s} & \frac{4 \sqrt{2} \left( m_{\bar{c}} + m_s \right) \left( m_{\bar{s}} + m_c \right)}{\sqrt{3} m_{\bar{c}} m_{\bar{s}} m_c m_s} \\
0 &  - \frac{16}{m_{\bar{c}} m_s} - \frac{16}{ m_{\bar{s}} m_c} & \frac{4 \sqrt{2} \left( m_{\bar{c}} + m_s \right) \left( m_{\bar{s}} + m_c \right)}{\sqrt{3} m_{\bar{c}} m_{\bar{s}} m_c m_s} & 0 \\
\frac{8 \sqrt{2} \left( m_{\bar{c}} - m_s \right) \left( m_{\bar{s}} - m_c \right)}{3 m_{\bar{c}} m_{\bar{s}} m_c m_s} & \frac{4 \sqrt{2} \left( m_{\bar{c}} + m_s \right) \left( m_{\bar{s}} + m_c \right)}{\sqrt{3} m_{\bar{c}} m_{\bar{s}} m_c m_s} & \frac{-8 m_c m_s - \left( 28 m_c + 2 m_s + 8 m_{\bar{s}} \right)m_{\bar{c}} - \left( 2 m_c + 28 m_s \right)m_{\bar{s}}}{3 m_{\bar{c}} m_{\bar{s}} m_c m_s} & \frac{- 4 m_c m_s + 14 m_{\bar{c}} m_c + 14 m_{\bar{s}} m_s - 4 m_{\bar{c}}m_{\bar{s}}}{\sqrt{3} m_{\bar{c}} m_{\bar{s}} m_c m_s} \\
\frac{4 \sqrt{2} \left( m_{\bar{c}} + m_s \right) \left( m_{\bar{s}} + m_c \right)}{\sqrt{3} m_{\bar{c}} m_{\bar{s}} m_c m_s} & 0 & \frac{-4 m_c m_s + 14 m_{\bar{c}} m_c + 14 m_{\bar{s}} m_s - 4 m_{\bar{c}}m_{\bar{s}}}{\sqrt{3} m_{\bar{c}} m_{\bar{s}} m_c m_s} & \frac{2}{m_{\bar{c}} m_s} + \frac{2}{m_{\bar{s}} m_c}
\end{array}\right),
\label{K-factor}
\end{align}
\end{widetext}
where the matrix is expressed in the CS bases given in Eq.~(\ref{CSbases2}). The quark masses are presented separately for antiquarks and quarks.
By subtracting the threshold term $-\frac{16}{m_{\bar{c}} m_s} - \frac{16}{m_{\bar{s}} m_c}$ from the diagonal elements, one can obtain the effective attraction coming from the color-spin interaction factor, $K_{X(3915)}-K_{D_s}-K_{\bar{D}_s}$.

For a tetraquark configuration, there are four CS bases for total spin 0 and six CS bases for total spin 1\cite{Park:2018wjk}.
For the $X(3915)$ with total spin 0, all four CS bases satisfy the charge conjugation symmetry, whereas in tetraquarks with similar structures but total spin 1, such as the $X(3872)$, this is not the case.
In fact, as noted in Ref.~\cite{Yun:2022evm}, only two of the six CS bases satisfy the quantum number constraints of the $X(3872)$, corresponding to the color singlet and color octet bases with the same spin structures.
This leads to mixing terms between color singlet and color octet bases for the $X(3915)$, as reflected in the non-zero off-diagonal elements of Eq.~(\ref{K-factor}).

When analyzing the diagonal elements of Eq.~(\ref{K-factor}), one finds that the expectation value corresponding to the $\left|CS_3\right\rangle$ state has a more intricate structure compared to those of the $\left|CS_1\right\rangle$, $\left|CS_2\right\rangle$, and $\left|CS_4\right\rangle$ basis states.
This difference arises because, within color SU(3) and spin SU(2), the color-color and spin-spin expectation values vanish for quark pairs directly coupled in singlet-singlet configurations of either the color or spin state.
Consequently, for the $\left|CS_1\right\rangle$, $\left|CS_2\right\rangle$, and $\left|CS_4\right\rangle$ bases, only the $(\bar{c}s)$ and $(\bar{s}c)$ pairs contribute.
In contrast, the $\left|CS_3\right\rangle$ basis corresponds to a color octet-octet and spin triplet-triplet configuration, for which the expectation values do not vanish, leading to nonzero contributions from all possible quark pairs.

On the other hand, from the diagonal elements of Eq.~(\ref{K-factor}), one might naively expect that the dominant contribution to the $X(3915)$ arises from the CS basis $|CS_3\rangle$, corresponding to the color octet and spin 1 states for both the $(\bar{c}s)$ and $(\bar{s}c)$ pairs.
This is based on the fact that, as shown from the (3,3) component of Eq.~(\ref{K-factor}), there exists an attractive contribution coming from the interaction of the $s\bar{s}$ pair, which is not suppressed by the relatively large charm quark mass $m_c$.
This attraction comes from the $|CS'_1\rangle$ , which corresponds to the color singlet and spin 0 for the $s\bar{s}$ pair contained in the $|CS_3\rangle$ state, as shown in Eq.~(\ref{CSbases2}).

However, once the full hyperfine interaction is evaluated by incorporating the spatial wave functions, the actual dominant attractive contribution is found to come from the CS basis $|CS_2\rangle$, which corresponds to the color singlet and spin 0 states for both the $(\bar{c}s)$ and $(\bar{s}c)$ pairs, i.e., the $D_s \bar{D}_s$ threshold configuration.
This difference arises because the actual contribution of the hyperfine potential depends not only on the color-spin factors but also on the spatial factor, which depends on the sizes of the individual quark pairs and, therefore, varies from pair to pair.
In particular, when the $D_s \bar{D}_s$ threshold mesons are largely separated, the distance between the $s$ and $\bar{s}$ quarks becomes large, leading to a suppression of their hyperfine interaction despite the favorable color-spin factor.
By combining the spatial factors associated with the variational parameters determined for the ground state of the $X(3915)$ and using the model parameters given in Eq.~(\ref{SimpleFitParameters}), the hyperfine potential matrix in Eq.~(\ref{Hamiltonian}) is evaluated in MeV unit as follows:
\begin{align}
\left\langle H^{Hyp} \right\rangle
=
\left(
\begin{array}{cccc}
74.99  & 0 & -0.41 & 0.36 \\
0 &  - 224.95 & 0.36 & 0 \\
-0.41 & 0.36 & - 10.39& 0.88 \\
0.36 & 0 & 0.88 & 28.12
\end{array}\right),
\label{HyperfineM}
\end{align}
where $H^{Hyp} = -\frac{3}{4}\sum^4_{i<j} \frac{\lambda^{c}_{i}}{2} \,\, \frac{\lambda^{c}_{j}}{2} V^{CS}_{ij}$.
As shown in Eq.~(\ref{HyperfineM}), the contribution from the $|CS_2\rangle$ is clearly dominant. 
This is further supported by the probability amplitudes listed in Table~\ref{Probability}, which shows the contributions of each CS basis to the ground state wave function of the $X(3915)$, obtained through the variational method.
From Table~\ref{Probability}, it is also evident that the ground state wave function of the $X(3915)$ expanded by the CS basis set of Eq.~(\ref{CSbases2}) corresponds to a well-separated $D_s \bar{D}_s$ configuration.
%%%%%%%%%%%%%%%%%%%%%%%%%%%		Table 4. Probability Amplitudes		%%%%%%%%%%%%%%%%%%%%%%%%%%%%%%%%%%%%%%
\begin{table}[h]

\caption{Probability amplitudes for each color-spin state in the ground state wave function of the $X(3915)$. The sets of color-spin bases in the table are defined in Eqs.~(\ref{CSbases}) and (\ref{CSbases2}).}

\centering

\begin{tabular}{ccc}
\hline
\hline
CS basis set	& Amplitudes \\
\hline 
$\left\{ \left|CS_1\right\rangle, \left|CS_2\right\rangle, \left|CS_3\right\rangle, \left|CS_4\right\rangle \right\}$	&	$\left\{ 0.0, 1.0, 0.0, 0.0 \right\}$		\\
$\left\{ \left|CS'_1\right\rangle, \left|CS'_2\right\rangle, \left|CS'_3\right\rangle, \left|CS'_4\right\rangle \right\}$	&	$\left\{ 0.03, 0.22, 0.08, 0.67 \right\}$	\\
\hline 
\hline
\label{Probability}
\end{tabular}
\end{table}
%%%%%%%%%%%%%%%%%%%%%%%%%%%%%%%%%%%%%%%%%%%%%%%%%%%%%%%%%%%%%%%%%%%%%%%%%%%%%%%%%%%%%%%%

As shown in Table~\ref{Probability}, the probability amplitudes obtained in a different basis set provide additional insight on the internal structure of the $X(3915)$ that the relative contributions of each quark pair in the tetraquark configuration of the $X(3915)$.
In particular, from Eq.~(\ref{CSbases2}), the transformation of the $|CS_2\rangle$ shows that the probability amplitudes obtained in the $(\bar{c}c) \otimes (\bar{s}s)$ basis are equivalent to the square of coefficients of each $(\bar{c}c) \otimes (\bar{s}s)$ basis.
In this transformation, the CS basis $|CS'_4\rangle$ has the largest probability amplitude.
This basis, in which both the $(\bar{c}c)$ and $(\bar{s}s)$ pairs are in color octet and spin 1 configurations, is structurally analogous to the dominant contributing basis for the $X(3872)$, which has a similar configuration but total spin 1.

Furthermore, considering a well-separated $D_s \bar{D}_s$ configuration, one can ignore the $1/(m_{\bar{c}} m_c)$, $1/(m_{\bar{s}} m_s)$, $1/(m_{\bar{c}} m_{\bar{s}})$, and $1/(m_c m_s)$ terms in Eq.~(\ref{K-factor}), corresponding to $(\bar{c}c)$, $(\bar{s}s)$, $(\bar{c}\bar{s})$, and $(cs)$ pairs, respectively.
In this limit, Eq.~(\ref{K-factor}) reduces to the following diagonal matrix.
\begin{align}
K_{X(3915)}=
\small
\left(
\begin{array}{cccc}
\frac{32}{3 m_c m_s}  & 0 & 0 & 0 \\
0 &  - \frac{32}{m_c m_s} & 0 & 0 \\
0 & 0 & - \frac{4}{3 m_c m_s}& 0 \\
0 & 0 & 0 & \frac{4}{m_c m_s}
\end{array}\right),
\label{reduced_K-factor}
\end{align}
where we take only the contributions corresponding to the threshold mesons and $m_{\bar{q}}=m_q$.
This further indicates that the ground state we calculate corresponds to a well-separated $D_s \bar{D}_s$ configuration.
The off-diagonal elements in Eq.~(\ref{HyperfineM}) vanish as the variational parameter $a_3 \to 0$, corresponding to this well-separated $D_s \bar{D}_s$ configuration, and the ratios of the diagonal elements reduce to those in Eq.~(\ref{reduced_K-factor}).
This suggests that the $X(3915)$ is unlikely to correspond to a compact configuration and is instead more consistent with a molecular structure, originating from the long-range interactions that extend beyond the scope of this model.
Nevertheless, although we find that the ground state favors a separated $D_s \bar{D}_s$ configuration, the model has an intrinsic uncertainty on the order of a few MeV, as shown in Table~\ref{FitResult}.
Consequently, a compact tetraquark structure cannot be conclusively ruled out.

In the following section, we study the possibility of experimentally discriminating among different candidate structures of the $X(3915)$ by analyzing additional physical observables associated with its production in heavy ion collisions within the framework of the coalescence model.

\section{Production of the $X(3915)$ meson in heavy ion collisions}

We consider here $X(3915)$ mesons produced from various
constituents by coalescence in relativistic heavy ion collisions.
Adopting the coalescence picture for the formation of the meson
\cite{Greco:2003xt, Greco:2003mm}, we evaluate the yield and
transverse momentum distribution of the $X(3915)$ in a $c\bar{c}$
charmonium, a $c\bar{c}s\bar{s}$ tetraquark, or a $D_s\bar{D}_s$
hadronic molecular state, denoted by $X_{c\bar{c}}$,
$X_{c\bar{c}s\bar{s}}$, and $X_{D_s\bar{D}_s}$, respectively.

\subsection{The $X(3915)$ meson in a two-quark state}

As mentioned before, we consider the $X(3915)$ as a $c\bar{c}$
state with an internal relative momentum, i.e., a $2p$-wave state.
When the $X(3915)$ meson is produced from one charm and one
anti-charm quarks in the quark-gluon plasma by charm quark
recombination, the production process of the $X(3915)$ meson is
similar to those of the $J/\psi$, $\chi_{c1}(1P)$, and $\psi(2S)$
mesons since the $X(3915)$ has the same quark contents as the above
charmonium states except for its different $2p$ internal structure.
Then, the transverse momentum distribution of the $X(3915)$ is
given by \cite{Cho:2014xha, Cho:2023kpe},
\begin{eqnarray}
&& \frac{d^2N_{X_{c\bar{c}}}}{d^2\vec p_T}=\frac{g_X}{V} \int
d^3\vec r d^2\vec p_{\bar{c}T}d^2\vec p_{cT} W_{2p}(\vec r,
\vec k) \nonumber \\
&& \qquad\qquad\times\frac{d^2N_{\bar{c}}}{d^2 \vec p_{\bar{c}T}}
\frac{d^2N_{c}} {d^2\vec p_{cT}}\delta^{(2)}(\vec p_T-\vec
p_{\bar{c}T} -\vec p_{cT}), \label{CoalTransX3915ccbar}
\end{eqnarray}
in the non-relativistic limit. In Eq. (\ref{CoalTransX3915ccbar})
the assumption on the boost-invariant longitudinal momentum
distributions for constituents satisfying $\eta=y$ between
spatial, $\eta$ and momentum $y$ rapidities, or the Bjorken
correlation has been made. The $g_X$ is the degeneracy factor for
possible chances of producing the $X(3915)$ meson from its
constituents, e.g., $g_X=1/(2\cdot 3)^2$.

The $\vec r$ and $\vec k$ in Eq. (\ref{CoalTransX3915ccbar}) are a
relative distance and a relative transverse momentum between charm
and anti-charm quarks, respectively, in the rest frame of the
$X(3915)$. We take here the following configurations and relative
transverse momenta for charm and anti-charm quarks,
\begin{eqnarray}
&& \vec R=\frac{m_{\bar{c}}\vec r_{\bar{c}}+m_c\vec r_c}
{m_{\bar{c}}+m_c}, \qquad \vec r=\vec r_{\bar{c}}-\vec r_c, \nonumber \\
&& \vec K=\vec p_{\bar{c}T}'+\vec p_{cT}', \qquad \vec
k=\frac{m_c\vec p_{\bar{c}T}'-m_{\bar{c}}\vec
p_{cT}'}{m_{\bar{c}}+m_c}. \label{rel_coordinates}
\end{eqnarray}

The relative transverse momentum $\vec k$ connects transverse
momenta of constituent quarks in two different frames, i.e.,
between $\vec p_{\bar{c}T}$ and $\vec p_{cT}$ in the fireball
frame and $\vec p_{\bar{c}T}'$ and $\vec p_{cT}'$ in the $X(3915)$
rest frame by Lorentz transformation \cite{Scheibl:1998tk,
Oh:2009zj}. The transverse momentum distribution difference
between the $X(3915)$ and other charmonium states in two-quark
states in the coalescence model originates from this Lorentz
transformation, as the mass difference between the $X(3915)$ and
other charmonium states causes different Lorentz transformation
from the fireball frame to the rest frame of the meson.

In regard to the $2p$-wave Wigner function, we construct it based
on harmonic oscillator wave functions in three dimensions,
\begin{equation}
\psi_{11m}(\vec r)=\frac{1}{(\pi\sigma^2)^{1/4}}\frac{4}
{\sqrt{15}}\frac{r}{\sigma^2}\Big(\frac{5}{2}-\frac{r^2}{\sigma^2}\Big)
e^{-\frac{r^2}{2\sigma^2}}Y_{1m}(\theta,\phi), \label{psir11m}
\end{equation}
with $\sigma^2=1/(\mu\omega)$ relating the oscillator frequency of
the harmonic oscillator wave function, $\omega$ and the reduced
mass, $\mu=m_{\bar{c}}m_c/(m_{\bar{c}}+m_c)$. In Eq.
(\ref{psir11m}), the subscript $11m$ in the wave function
$\psi_{11m}(\vec r)$ represents the quantum number $klm$ in the
three dimensional harmonic oscillator, i.e., the energy
$E_n=(n+3/2)\hbar\omega$ with $n=2k+l$, and $Y_{lm}(\theta,\phi)$
is the spherical harmonics.

Using the $m$-averaged density for $2p$ states, similar to that
for $1d$ states \cite{Cho:2011ew},
\begin{equation}
\rho(\vec r,\vec r')=\frac{1}{3}\sum_m \psi_{11m}(\vec
r)\psi_{11m}^*(\vec r')
\end{equation}
we build the Wigner function,
\begin{equation}
W(\vec r, \vec k)=\int d^3\vec q \rho\Big(\vec r+\frac{\vec
q}{2},\vec r-\frac{\vec q}{2}\Big)e^{i\vec k\cdot\vec q}
\label{Wigner}
\end{equation}
resulting in,
\begin{eqnarray}
&& W_{2p}(\vec r, \vec k)=\frac{15}{32}e^{-\frac{r^2}{\sigma^2}
-k^2\sigma^2}\bigg(\frac{r^6}{\sigma^6}-\frac{11}{2}\frac{r^4}{\sigma^4}+
\frac{15}{2}\frac{r^2}{\sigma^2}+\frac{15}{2}k^2\sigma^2 \nonumber \\
&& \qquad\qquad-\frac{11}{2}k^4\sigma^4+k^6\sigma^6+r^2k^2\Big(3-
\frac{r^2}{\sigma^2}-k^2\sigma^2\Big) \nonumber \\
&& \qquad\qquad +2(\vec r\cdot\vec k)\Big(-7+2\frac{r^2}{\sigma^2}
+2k^2\sigma^2\Big)-\frac{15}{4}\bigg). \label{Winger2p}
\end{eqnarray}

The $2p$-Wigner function in Eq. (\ref{Winger2p}) is consistent
with the previous result \cite{Kordell:2021prk}, and satisfies,
\begin{equation}
\int d^3\vec rW_{2p}(\vec r,\vec k)=\frac{1}{3}\sum_m
\tilde{\psi}_{11m}(\vec k)\tilde{\psi}_{11m}^*(\vec k)
\label{WigIntr}
\end{equation}
with the three dimensional harmonic oscillator wave function in
momentum space, $\tilde{\psi}_{11m}(\vec k)$.

Adopting the above $2p$-Wigner function in Eq. (\ref{Winger2p}) to
the transverse momentum distribution of the $X(3915)$ in a $2p$
charmonium state in Eq. (\ref{CoalTransX3915ccbar}) yields,
\begin{eqnarray}
&& \frac{d^2N_{X_{c\bar{c}}}}{d^2\vec p_T}=\frac{g_X}{V}
(2\sqrt{\pi}\sigma)^3 \int d^2\vec p_{\bar{c}T}d^2\vec p_{cT}
\frac{d^2N_{\bar{c}}}{d^2 \vec p_{\bar{c}T}} \frac{d^2N_{c}}
{d^2\vec p_{cT}} \nonumber \\
&& \qquad\quad\times\frac{4}{15}\sigma^2k^2e^{-\sigma^2
k^2}\Big(k^2\sigma^2-\frac{5}{2}\Big)^2\delta^{(2)}(\vec p_T-\vec
p_{\bar{c}T} -\vec p_{cT}). \nonumber \\
\label{CoalTrans2qX3915}
\end{eqnarray}

\subsection{The $X(3915)$ meson in a four-quark state}

Next, we calculate the transverse momentum distribution of the
$X(3915)$ meson when it exists in a four-quark state: the
$X(3915)$ meson is produced from one strange, one anti-strange,
one charm and one anti-charm quarks in the quark-gluon plasma by
coalescence. All quarks inside the $X(3915)$ in a four-quark state
reside in a $s$-wave, consistent with its spin and parity,
$J^P=0^+$.

The transverse momentum distribution of the $X(3915)$ meson in a
four-quark state is taken from that of the $X(3872)$ meson in a
four-quark state \cite{Cho:2019syk}, replacing two light quarks
with two strange quarks,

\begin{eqnarray}
&& \frac{d^2N_{X_{c\bar{c}s\bar{s}}}}{d^2\vec
p_T}=\frac{g_X}{V^3}(2\sqrt{\pi})^9 (\sigma_1\sigma_2\sigma_3)^3
\nonumber \\
&& \qquad\qquad\quad \times\int d^2\vec p_{sT}d^2\vec
p_{\bar{s}T}d^2\vec p_{cT}d^2\vec p_{\bar{c}T} \nonumber \\
&& \qquad\qquad\quad \times \frac{d^2N_s}{d^2 \vec p_{sT}}
\frac{d^2 N_{\bar{s}}}{d^2\vec p_{\bar{s}T}} \frac{d^2N_c}{d^2\vec
p_{cT}} \frac{d^2N_{\bar{c}}}{d^2\vec p_{\bar{c}T}} \nonumber \\
&& \qquad\qquad\quad \times\delta^{(2)}(\vec p_T-\vec p_{sT}-\vec
p_{\bar{s}T}-\vec p_{cT}-\vec p_{\bar{c}T}) \nonumber \\
&& \qquad\qquad\quad \times \exp{\bigg(-\sigma_1^2
k_1^2-\sigma_2^2k_2^2-\sigma_3^2k_3^2\bigg)},
\label{CoalTransX3915sscc}
\end{eqnarray}
with relative coordinates,

\begin{eqnarray}
&& \vec R_{4q}=\frac{m_s\vec r_s+m_{\bar{s}}\vec
r_{\bar{s}}+m_c\vec r_c+m_{\bar{c}}\vec
r_{\bar{c}}}{m_s+m_{\bar{s}}+m_c+m_{\bar{c}}},
\nonumber \\
&& \vec r_1=\vec r_s-\vec r_{\bar{s}}, \nonumber \\
&& \vec r_2=\frac{m_s\vec r_s+m_{\bar{s}}\vec
r_{\bar{s}}}{m_s+m_{\bar{s}}}-\vec r_c, \nonumber \\
&& \vec r_3=\frac{m_s\vec r_s+m_{\bar{s}}\vec r_{\bar{s}}+m_c\vec
r_c}{m_s+m_{\bar{s}}+m_c}-\vec r_{\bar{c}}, \label{rel_coord}
\end{eqnarray}
and corresponding relative transverse momenta,

\begin{eqnarray}
&& \vec k_{4q}=\vec p_{sT}'+\vec p_{\bar{s}T}'+\vec p_{cT}'+\vec
p_{\bar{c}T}', \nonumber \\
&& \vec k_1=\frac{m_{\bar{s}}\vec p_{sT}'-m_s\vec
p_{\bar{s}T}'}{m_s+m_{\bar{s}}}, \nonumber \\
&& \vec k_2=\frac{m_c(\vec p_{sT}'+\vec
p_{\bar{s}T}')-(m_s+m_{\bar{s}})
\vec p_{cT}'}{m_s+m_{\bar{s}}+m_c}, \nonumber \\
&& \vec k_3=\frac{m_{\bar{c}}(\vec p_{sT}'+\vec p_{\bar{s}T}'+\vec
p_{cT}')-(m_s+m_{\bar{s}}+m_c)\vec
p_{\bar{c}T}'}{m_s+m_{\bar{s}}+m_c
+m_{\bar{c}}}. \nonumber \\
\label{rel_moment}
\end{eqnarray}

In Eq. (\ref{rel_moment}), the transverse momenta of quarks, $\vec
p_{qT}'$ ($q=s, \bar{s}, c, \bar{c}$) are the transverse momenta
in the rest frame of the $X(3915)$ meson, obtained from those in
the fireball frame, $\vec p_q$ by Lorentz transformation as
mentioned in the previous section. The $\sigma_1$, $\sigma_2$ and
$\sigma_3$ in Eq. (\ref{CoalTransX3915sscc}) are related to the
oscillator frequency of the harmonic oscillator wave function
$\omega$ in the $s$-wave Wigner function,

\begin{eqnarray}
&& W_{c\bar{c}s\bar{s}}(\vec r_1, \vec r_2, \vec r_3, \vec
k_1, \vec k_2, \vec k_3) \nonumber \\
&& =8^3\exp{\bigg(-\frac{r_1^2}{\sigma_1^2}-\sigma_1^2k_1^2
\bigg)}\exp{\bigg(-\frac{r_2^2}{\sigma_2^2}-\sigma_2^2k_2^2\bigg)}
\nonumber \\
&& \times\exp{\bigg(-\frac{r_3^2}{\sigma_3^2}-\sigma_3^2
k_3^2\bigg)}, \label{wigner4q}
\end{eqnarray}
with reduced masses,

\begin{eqnarray}
&& \mu_1=\frac{m_sm_{\bar{s}}}{m_s+m_{\bar{s}}}, \quad \mu_2=
\frac{(m_s+m_{\bar{s}})m_c}{m_s+m_{\bar{s}}+m_c}, \nonumber \\
&& \mu_3=\frac{(m_s+m_{\bar{s}}+m_c)m_{\bar{c}}}{m_s+m_{\bar{s}}+
m_c+m_{\bar{c}}},
\end{eqnarray}
corresponding to $\sigma_1^2=1/(\omega\mu_1)$,
$\sigma_2^2=1/(\omega\mu_2)$, and $\sigma_3^2=1/(\omega\mu_3)$,
respectively.

In taking the relative configurations between four quarks, one may
choose other coordinates and momenta rather than Eqs.
(\ref{rel_coord}) and (\ref{rel_moment}). However, it has been
shown that the yield and transverse momentum distribution are not
dependent on different choices for relative coordinates and
momenta between quarks inside the hadron when all quarks are in
the ground state, suitable for the $s$-wave Gaussian Wigner
function, Eq. (\ref{wigner4q}) with one oscillator frequency
$\omega$ \cite{Cho:2019syk}. We take the oscillator frequency,
0.103 GeV \cite{Cho:2025lrc} in Eqs. (\ref{CoalTransX3915ccbar})
and (\ref{CoalTransX3915sscc}).

\subsection{The $X(3915)$ meson in a hadronic molecular state}

When the $X(3915)$ meson is in a hadronic molecular state made up
of $D_s$ and $\bar{D}_s$ mesons, the yield and transverse momentum
distribution of it can be obtained by replacing charm and
anti-charm quarks with $D_s$ and $\bar{D}_s$ mesons, respectively
in Eq. (\ref{CoalTransX3915ccbar}). Thus, the transverse momentum
distribution of the $X_{D_s\bar{D}_s}$ is given by,

\begin{eqnarray}
&& \frac{d^2N_{X_{D_s\bar{D}_s}}}{d^2\vec
p_T}=\frac{g_X}{V}(2\sqrt{\pi}\sigma_h)^3\int d^2\vec
p_{D_sT}d^2\vec
p_{\bar{D}_sT} e^{-\sigma_h^2 k_h^2} \nonumber \\
&& \qquad\qquad\quad\times\frac{d^2N_{D_s}}{d^2 \vec p_{D_sT}}
\frac{d^2N_{\bar{D}_s}} {d^2\vec p_{\bar{D}_sT}}\delta^{(2)}(\vec
p_T-\vec p_{\bar{D}_sT} -\vec p_{D_{s}T}), \nonumber \\
\label{CoalTransX3915DsDbars}
\end{eqnarray}
with relative coordinates and transverse momenta between $D_s$ and
$\bar{D}_s$ mesons,

\begin{eqnarray}
&& \vec R_h=\frac{m_{\bar{D}_s}\vec r_{\bar{D}_s}+m_{D_s}\vec
r_{D_s}}{m_{\bar{D}_s}+m_{D_s}}, \qquad \vec r_h=\vec r_{\bar{D}_s}-
\vec r_{D_s}, \nonumber \\
&& \vec K_h=\vec p_{\bar{D}_{s}T}'+\vec p_{D_{s}T}', \qquad \vec
k_h=\frac{m_{D_s}\vec p_{\bar{D}_{s}T}'-m_{\bar{D}_s}\vec
p_{D_{s}T}'}{m_{\bar{D}_s}+m_{D_s}}, \nonumber \\
\label{rel_coordinatesDsDbars}
\end{eqnarray}
again with the relative transverse momentum $\vec k_h$ connecting
transverse momenta of $D_s$ and $\bar{D}_s$ mesons between $\vec
p_{\bar{D}_{sT}}$ and $\vec p_{D_{sT}}$ in the fireball frame and
$\vec p_{\bar{D}_{sT}}'$ and $\vec p_{D_{sT}}'$ in the rest frame
of the $X(3915)$ in a hadronic molecular state by Lorentz
transformation.

In Eq. (\ref{rel_coordinatesDsDbars}), the $s$-wave Wigner
function,

\begin{equation}
W_s(\vec r, \vec k) = 8 e^{-\frac{r^2}{\sigma_h^2}-k^2 \sigma_h^2}
\label{Wigners}
\end{equation}
has already been adopted with both the $\sigma_h$, representing
the oscillator frequency of the harmonic wave function in the
Wigner function, and the reduced mass,
$\mu_h=m_{D_s}m_{\bar{D}_{s}}/(m_{D_s}+m_{\bar{D}_{s}})$. We take
the oscillator frequency, 0.0876 GeV obtained from the relation
between the binding energy and the oscillator frequency
\cite{Cho:2011ew}, $\omega=6 E_B$, with the binding energy of the
$X(3915)$, $E_B=M_{D_s}+M_{\bar{D}_{s}}-M_{X(3915)}=1968.35\times
2-3922.1=14.6$ (MeV).

\section{Yields and transverse momentum distributions of
the $X(3915)$}

We evaluate the transverse momentum distributions of $X(3915)$
mesons using Eqs. (\ref{CoalTrans2qX3915}),
(\ref{CoalTransX3915sscc}) and (\ref{CoalTransX3915DsDbars}) when
they are produced as either charmonium states, four-quark states,
or hadronic molecular states from their constituents by
recombination in heavy ion collisions at $\sqrt{s_{NN}}=5.02$ TeV.
As the production or the yield distribution of the $X(3915)$ as
functions of transverse momenta is dependent on its constituents,
it is essential to understand as a first step their transverse
momentum distributions in heavy ion collisions, especially at
$\sqrt{s_{NN}}=5.02$ TeV before calculating the yield or the
transverse momentum distribution of the $X(3915)$.

If the $X(3915)$ is in a two- or four-quark state, the $X(3915)$
meson can be regarded to be produced directly from charm and
strange quarks during the quark-hadron phase transition, thereby
it is necessary to know the transverse momentum distribution of
charm and strange quarks in the quark-gluon plasma. On the other
hand, if the $X(3915)$ is in a hadronic molecular state, the
information on the transverse momentum distribution of the $D_s$
and $\bar{D}_s$ mesons is required.

In order to describe the production of the $X(3915)$ in two- and
four-quark states, we adopt here the transverse momentum
distributions of charm and strange quarks at mid-rapidities in
central collisions obtained from the analysis on the production of
$\phi$ and $D^0$ mesons at $\sqrt{s_{NN}}=5.02$ TeV in
relativistic heavy ion collisions \cite{Cho:2025lrc},

\begin{widetext}
\begin{eqnarray}
&& \frac{d^2N_c}{d^2\vec p_{cT}}=\left\{
\begin{array}{ll}
1.63 ~\mathrm{(GeV^{-2})} e^{-0.27(p_{cT}/p_{0T})^{2.03}}, & \quad p_{cT} \le 1.80~\textrm{GeV} \\
7.95 ~\mathrm{(GeV^{-2})} e^{-3.49(p_{cT}/p_{0T})^{3.59}}
+\frac{90112
~\mathrm{(GeV^{-2})}}{(1.0+(p_{cT}/p_{0T})^{0.50})^{14.19}}, &
\quad p_{cT} > 1.80~\textrm{GeV}
\end{array} \right.
\end{eqnarray}
and
\begin{eqnarray} && \frac{d^2N_s}{d^2 \vec p_{sT}}=\left\{
\begin{array}{ll}
\frac{V}{(2\pi)^3}m_Te^{-m_T/T_{eff}}, & \quad p_{sT} \le 1.50~\textrm{GeV} \\
21.95 ~\mathrm{(GeV^{-2})} e^{-0.17(p_{sT}/p_{0T})^{3.23}}
+\frac{80112 ~\mathrm{(GeV^{-2})}}
{(1.0+(p_{sT}/p_{0T})^{0.65})^{10.29}}. & \quad p_{sT} >
1.50~\textrm{GeV}
\end{array} \right.
\label{d2NcsdpT2}
\end{eqnarray}
\end{widetext}

In Eq. (\ref{d2NcsdpT2}), $g_s$, $m_T=\sqrt{p_{T}^2+m^2}$, $V$ and
$T_{eff}$ are the degeneracy factor for the color and spin of
strange quarks, the transverse mass, the coalescence volume and
the effective temperature, respectively. The coalescence volume of
3360 fm$^3$ and the effective temperature of $T_{eff}$=173 MeV
have also been applied here \cite{Cho:2025lrc}. The $p_{0T}$ in
Eq. (\ref{d2NcsdpT2}) is taken as 1.0 GeV to make the arguments in
the transverse momentum distributions dimensionless.

In order to calculate the transverse momentum distribution of the
$X(3915)$ meson in $D_s\bar{D}_s$ hadronic molecular states, we
adopt here the result on the analysis of the $D_s$ meson
production, or the transverse momentum distribution of the $D_s$
meson evaluated from the above charm and strange quark transverse
momentum distributions \cite{Cho:2025lrc}. However, we consider
here two transverse momentum distributions of the $D_s$ meson; the
one at chemical freeze-out and the other at kinetic freeze-out.

It is certain that the hadronic molecular states are formed after
their constituents are produced at chemical freeze-out in heavy
ion collisions, but it is still uncertain exactly when they are
formed from their constituents between chemical and kinetic
freeze-outs in heavy ion collisions. In one extreme case, it is
possible for them to be produced right after their constituents
are produced at chemical freeze-out. In the other extreme case,
they can be produced at kinetic freeze-out after some interactions
between their constituents and light hadrons during the hadronic
stage. In between two cases, it is also possible that they are
produced anytime between chemical and kinetic freeze-outs.

Moreover, as hadrons in molecular states are mostly loosely bound,
they may be easily dissociated by light hadrons during the
hadronic stage. Thus, when those hadrons are produced either at
chemical freeze-out or anytime between two freeze-outs, their
yields are inevitably affected from their interactions between
hadrons until kinetic freeze-out, and therefore it is required to
investigate the hadronic effects on the production or dissociation
of hadrons in molecular states \cite{Cho:2013rpa, Torres:2014fxa,
Hong:2018mpk}. The yield, or the transverse momentum distributions
of hadrons in molecular states must be strongly dependent on when
they are formed in heavy ion collisions.

By this reason, we present here the yields and transverse momentum
distributions for two possibilities in the formation of the
$X(3915)$ meson from $D_s$ and $\bar{D}_s$ mesons in heavy
ion collisions. At first, we consider the formation of the
$X(3915)$ at chemical freeze-out from $D_s$ and $\bar{D}_s$ mesons
also produced at chemical freeze-out based on the transverse
momentum distribution of $D_s$ and $\bar{D}_s$ mesons evaluated in
the coalescence model \cite{Cho:2025lrc}. In this case, the
additional production of $D_s$ and $\bar{D}_s$ mesons from
feed-down contributions are overlooked due to a short time for
heavier charm-strange mesons to decay into $D_s$ mesons before the
$X(3915)$ is formed.

\begin{figure}[!t]
\begin{center}
\includegraphics[width=0.50\textwidth]{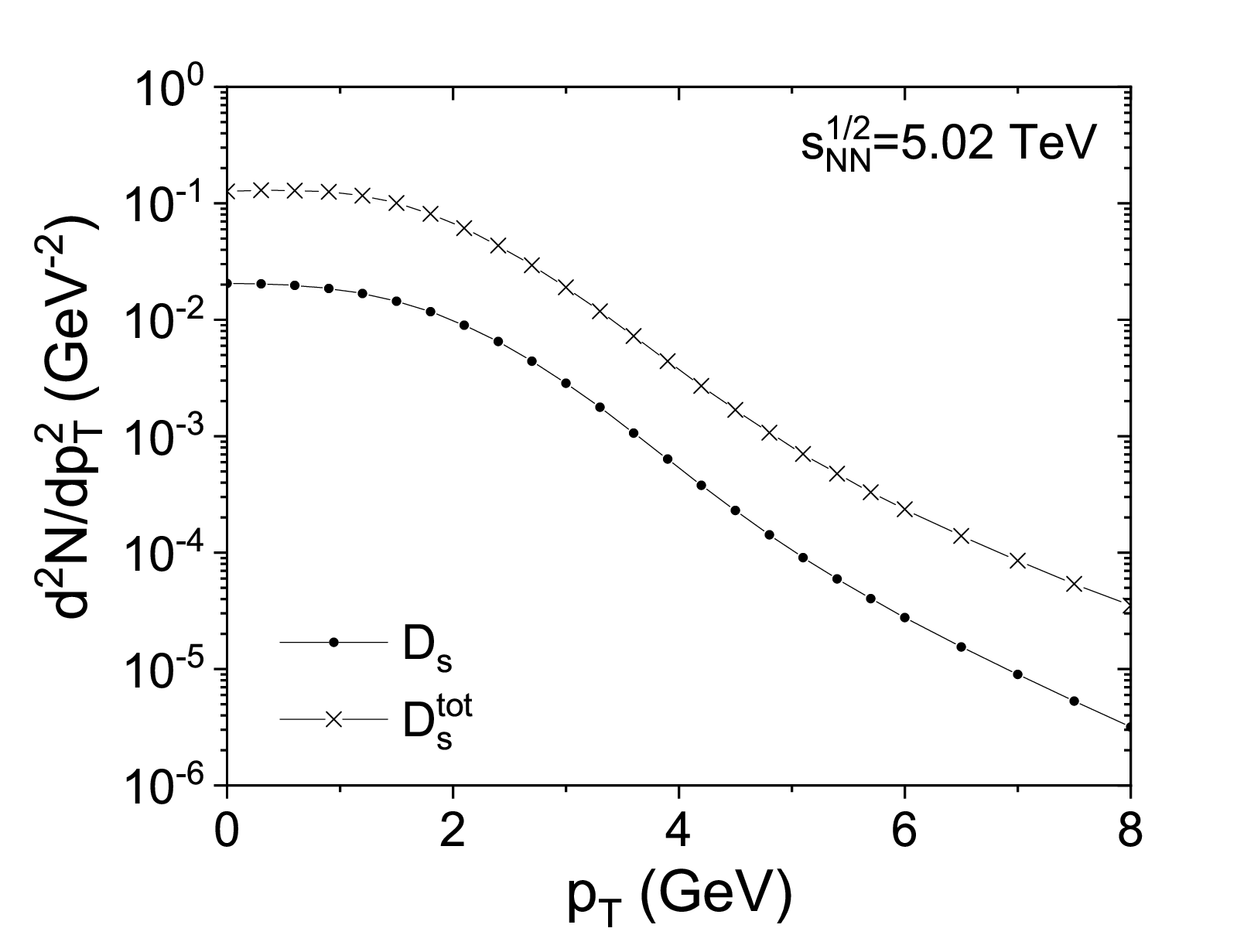}
\end{center}
\caption{Transverse momentum distributions of the $D_s$ meson,
$dN_{D_s}/dp_T$ at mid-rapidity at $\sqrt{s_{NN}}=5.02$ TeV with
and without feed down contributions, denoted by $D_s$ and
$D_s^{tot}$, respectively. } \label{DsDbars}
\end{figure}

At second, we adopt the transverse momentum distribution of $D_s$
and $\bar{D}_s$ mesons after considering the feed-down from
heavier charm-strange mesons, or $D_s^*$, $D_{s0}^*(2317)$, and
$D_{s1}(2460)$ mesons, also calculated in the coalescence model in order to take into account the formation of the $X(3915)$ at kinetic freeze-out \cite{Cho:2025lrc}: the yield and transverse momentum distribution of $D_s$ mesons
evaluated in the coalescence model including feed-down
contributions is shown to agree reasonably well with the measurement by ALICE
Collaboration \cite{Cho:2025lrc}. We show in Fig. \ref{DsDbars}
the transverse momentum distributions of $D_s$ mesons with and
without feed-down contributions.

With the above transverse momentum distribution of $D_s$ mesons
shown in Fig. \ref{DsDbars} as well as those of charm and strange
quarks, in Eq. (\ref{d2NcsdpT2}), we can calculate the transverse
momentum distribution of the $X(3915)$ meson for a hadronic
molecular state, a tetraquark state, and a charmonium state. We
take the coalescence volume of 3360 fm$^3$ \cite{Cho:2025lrc} for the
$X(3915)$ in a tetraquark or a charmonium state, and a hadronic
molecular state produced from bare $D_s$ mesons at chemical
freeze-out, while we adopt the volume at kinetic freeze-out, 52700 fm$^3$ determined by
requiring both the constant number of light particles and
conserving entropy per particle during the hadronic stage
\cite{Sung:2021myr} in order to describe the production of the
$X(3915)$ at kinetic freeze-out.

\begin{figure}[!t]
\begin{center}
\includegraphics[width=0.50\textwidth]{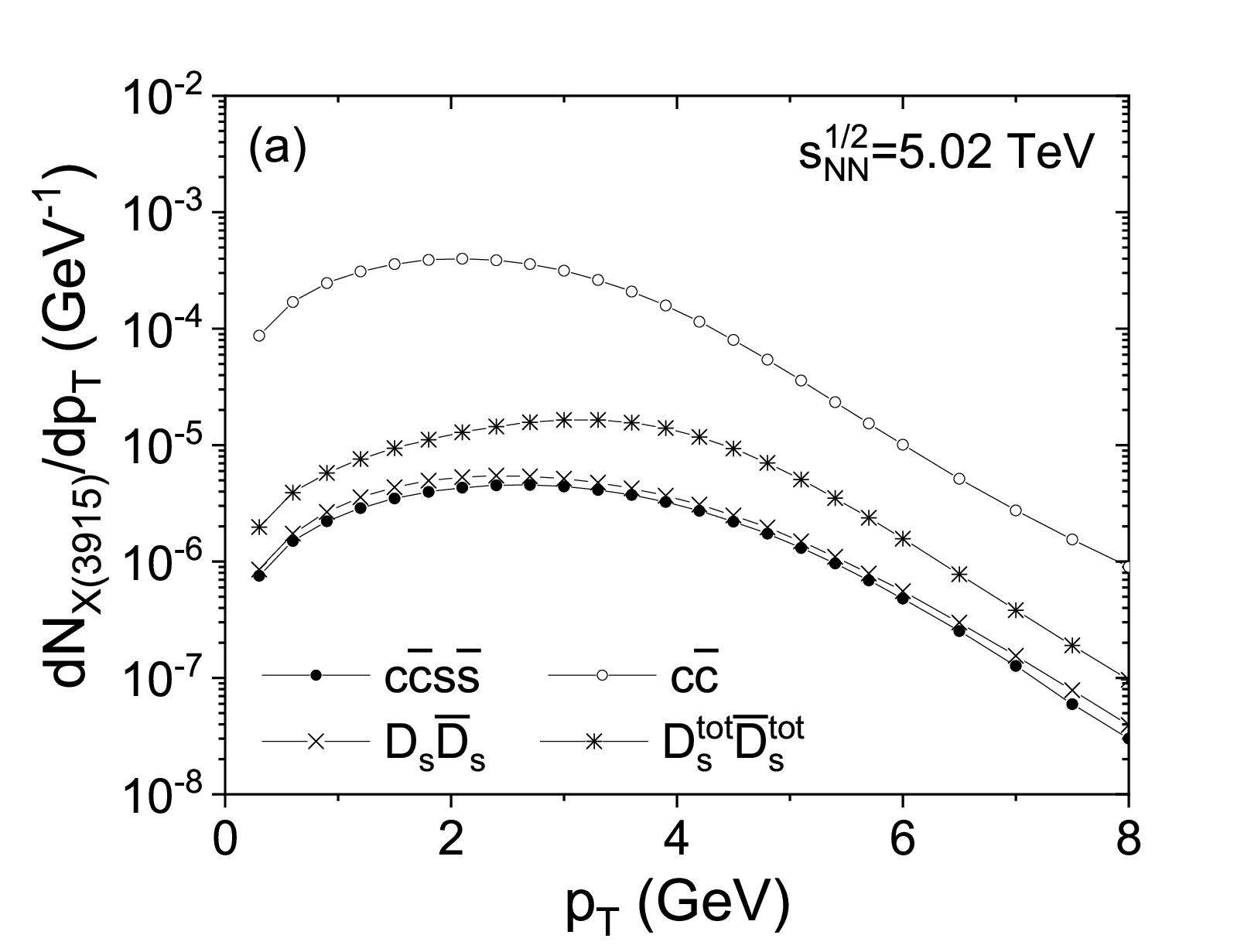}
\includegraphics[width=0.50\textwidth]{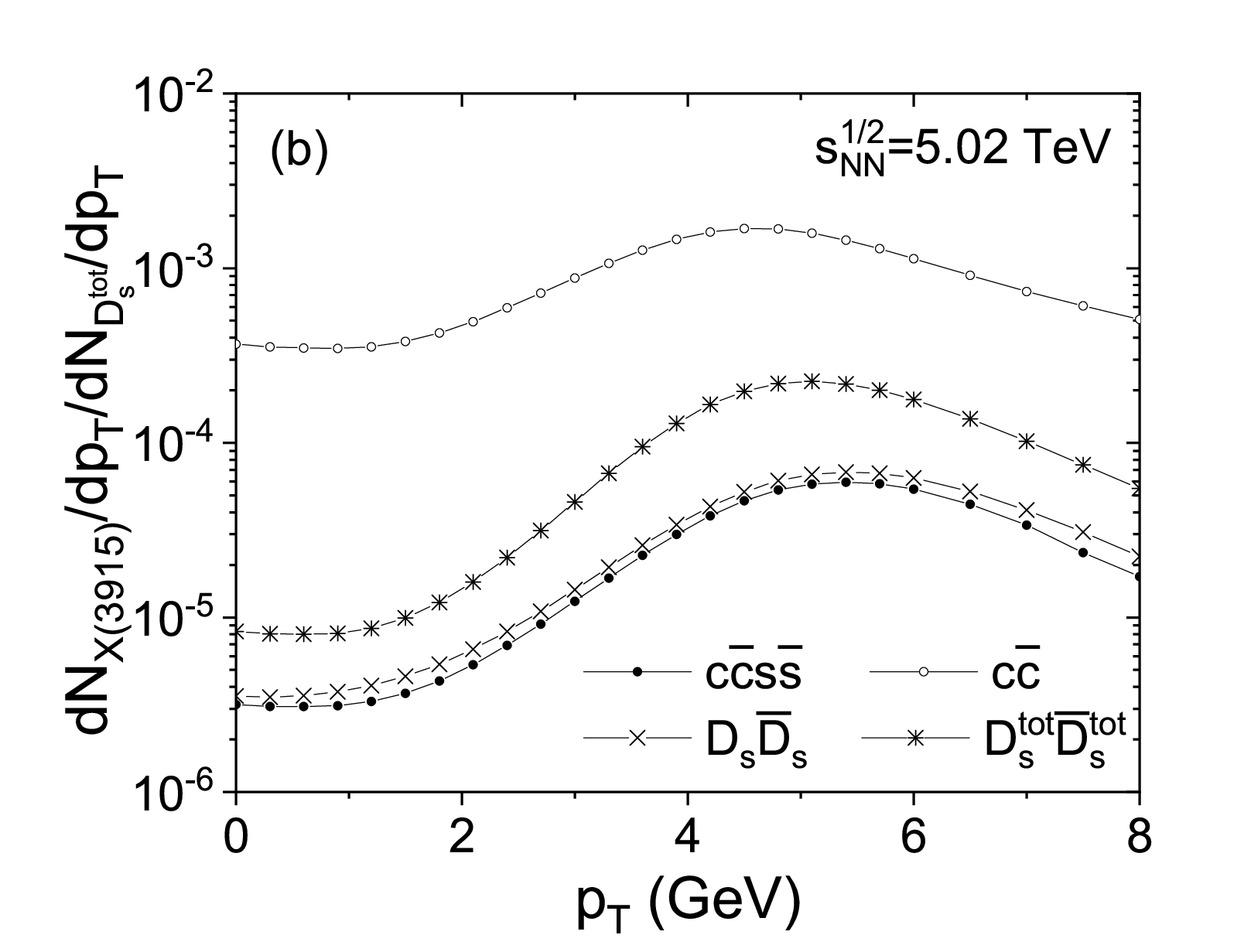}
\end{center}
\caption{(a) Transverse momentum distributions of the $X(3915)$
meson, $dN_{X(3915)}/dp_T$ at mid-rapidity at $\sqrt{s_{NN}}=5.02$
TeV for various possible states, or a charmonium state,
$c\bar{c}$, a tetraquark state, $c\bar{c}s\bar{s}$, a hadronic
molecular state formed from $D_s$ mesons at chemical freeze-out,
$D_s\bar{D}_s$, and a hadronic molecular state formed from $D_s$
mesons at kinetic freeze-out, $D_s^{tot}\bar{D}_s^{tot}$. (b)
Transverse momentum distribution ratios between the $X(3915)$ of
various states and the $D_s$ meson. } \label{dNdpTX3915}
\end{figure}

We show in Fig. \ref{dNdpTX3915}(a) transverse momentum
distributions of the $X(3915)$ meson, $dN_{X(3915)}/dp_T$ at
mid-rapidity in central collisions at $\sqrt{s_{NN}}=5.02$ TeV for
its possible states, or a charmonium state, $X_{c\bar{c}}$, a
tetraquark state, $X_{c\bar{c}s\bar{s}}$, and a hadronic molecular
state formed from $D_s$ mesons at chemical freeze-out,
$X_{D_s\bar{D}_s}$, and a hadronic molecular state formed from
$D_s$ mesons at kinetic freeze-out,
$X_{D_s^{tot}\bar{D}_s^{tot}}$. As shown in the Fig.
\ref{dNdpTX3915}(a) the transverse momentum distribution of the
$X(3915)$ in a normal charmonium state is highest, and that of
the $X(3915)$ in a hadronic molecular state formed from $D_s$ and
$\bar{D}_s$ mesons at kinetic freeze-out is next highest. Two
transverse momentum distribution of the $X(3915)$ both in a
tetraquark state and in a hadronic molecular state formed from
$D_s\bar{D}_s$ at chemical freeze-out are found to be very
similar.

We note that the formation of the $X(3915)$ from $D_s$ and
$\bar{D}_s$ mesons at chemical freeze-out without feed-down
contributions is almost the same as the production of the $X(3915)$ in
a tetraquark state when $c\bar{s}$ and $\bar{c}s$ are formed first,
followed by $c\bar{c}s\bar{s}$ formation, based on the following
relative coordinates,

\begin{eqnarray}
&& \vec R=\frac{m_s\vec r_s+m_{\bar{s}}\vec r_{\bar{s}}+m_c\vec
r_c+m_{\bar{c}}\vec r_{\bar{c}}}{m_s+m_{\bar{s}}+m_c+m_{\bar{c}}},
\nonumber \\
&& \vec r_1=\vec r_c-\vec r_{\bar{s}}, \nonumber \\
&& \vec r_2=\vec r_{\bar{c}}-\vec r_s, \nonumber \\
&& \vec r_3=\frac{m_c\vec r_c+m_{\bar{s}}\vec
r_{\bar{s}}}{m_c+m_{\bar{s}}}-\frac{m_{\bar{c}}\vec
r_{\bar{c}}+m_s\vec r_s}{m_{\bar{c}}+m_s}, \label{rel_coord2}
\end{eqnarray}
and corresponding relative transverse momenta,

\begin{eqnarray}
&& \vec k=\vec p_{sT}'+\vec p_{\bar{s}T}'+\vec p_{cT}'+\vec
p_{\bar{c}T}', \nonumber \\
&& \vec k_1=\frac{m_{\bar{s}}\vec p_{cT}'-m_c\vec
p_{\bar{c}T}'}{m_c+m_{\bar{s}}}, \nonumber \\
&& \vec k_2=\frac{m_s\vec p_{\bar{c}T}'-m_{\bar{c}}\vec
p_{sT}'}{m_{\bar{c}}+m_s}, \nonumber \\
&& \vec k_3=\frac{(m_{\bar{c}}+m_s)(\vec p_{cT}'+\vec
p_{\bar{s}T}')-(m_c+m_{\bar{s}})(\vec p_{\bar{c}T}'+\vec
p_{sT}')}{m_s+m_{\bar{s}}+m_c
+m_{\bar{c}}}. \nonumber \\
\label{rel_moment2}
\end{eqnarray}

As mentioned before, the yield or transverse momentum distribution
of the hadron in a tetraquark state is independent of different
configurations between quarks for their relative coordinates and
momenta, either Eqs. (\ref{rel_coord}) and (\ref{rel_moment}) or
Eqs. (\ref{rel_coord2}) and (\ref{rel_moment2}) on the condition
that all quarks are in $s$-wave states, suitable for the $s$-wave
Gaussian Wigner function with a common oscillator frequency
$\omega$ \cite{Cho:2019syk}. Thus, it is possible that the
transverse momentum distribution of the $X(3915)$ in a hadronic
molecular state formed from $D_s$ and $\bar{D}_s$ mesons at
chemical freeze-out can be almost same as that of the $X(3915)$ in
a tetraquark state. In that sense the yield or transverse momentum
distributions of the exotic hadrons in a multiquark state can play
a role as a limiting values for those of exotic hadrons in a
hadronic molecular state.

However, here in the case of the formation of the $X(3915)$ in a
hadronic molecular state, the different oscillator frequency
$\omega$=0.0876 GeV, representing for the loose binding between
$D_s$ and $\bar{D}_s$ mesons, has been adopted, thereby making the
above two transverse momentum distributions little bit different.
On the other hand, the transverse momentum distribution of the
$X(3915)$ in a hadronic molecular state produced at kinetic
freeze-out is evaluated to be about three times larger than that
of the $X(3915)$ produced at chemical freeze-out. The difference
between the transverse momentum distributions of the $X(3915)$ in
a harmonic molecular state at chemical and kinetic freeze-outs is
attributable to both the larger number of $D_s$ mesons at kinetic
freeze-out due to feed-down contributions and the larger
freeze-out volume, compared to the number of $D_s$ mesons with the
coalescence volume at chemical freeze-out; the number of $D_s$
mesons increases by 6.8 times while the volume of the system
increases by about 15.7 times between chemical and kinetic
freeze-out, thereby resulting in $6.8^2/15.7\approx 2.9$.

The different behaviors of transverse momentum distribution of the
$X(3915)$ meson in various possible states are shown more
explicitly when the ratios of the transverse momentum distribution
between the $X(3915)$ of those states and the $D_s$ meson are
taken. Moreover, when the decay of the $X(3915)$ meson to $D_s$
and $\bar{D}_s$ mesons has been considered
\cite{ParticleDataGroup:2024cfk}, the measurement of the ratio
between the $X(3915)$ and $D_s$ meson is expected to reduce
experimental uncertainties in the observation of the $X(3915)$.

We show in Fig. \ref{dNdpTX3915}(b) transverse momentum
distribution ratios between the $X(3915)$ of various states and
the $D_s$ meson. As shown in Fig. \ref{dNdpTX3915}(b), all the
ratios increase with increasing transverse momenta at low $p_T$, and have the
peak at the intermediate transverse momentum regions. The ratio
between the $X(3915)$ in a molecular state produced at kinetic
freeze-out and the $D_s$ meson increases by about 20 times in the
intermediate transverse momentum regions, while the ratio between
the $X(3915)$ in a charmonium state and the $D_s$ meson increases
by about 2 times, showing different behaviors in the ratios between
the $X(3915)$ and the $D_s$ depending on the structure of the
$X(3915)$.

\begin{table}[!t]
\caption{Yields of the $X(3915)$ for various possible states
evaluated from the integration of the transverse momentum
distributions shown in Fig. \ref{dNdpTX3915}(a) over all
transverse momenta at $\sqrt{s_{NN}}=5.02$ TeV at LHC. The yield
of the $X(3915)$ in a thermal model \cite{Cho:2025lrc} is also
listed for comparison.} \label{X3915yields}
\begin{center}
\begin{tabular}{c|c|c|c|c|c}
\hline \hline
& $c\bar{c}$ & $c\bar{c}s\bar{s}$ & $D_s\bar{D}_s$ & $D_s^{tot}\bar{D}_s^{tot}$ & Ther. \\
\hline Yields ($\times 10^{-4}$) & ~12.0~ & ~0.162~ & ~0.194~ & 0.567 & ~6.38~ \\
\hline \hline
\end{tabular}
\end{center}
\end{table}

Integrating transverse momentum distributions of the $X(3915)$
meson in its various states over the entire transverse momentum
region leads to their yields, and the results are shown in Table
\ref{X3915yields}. Also listed in Table \ref{X3915yields} is the
yield of the $X(3915)$ evaluated in a thermal model
\cite{Cho:2025lrc} for comparison. As argued on the transverse
momentum distribution of the $X(3915)$ before, the yield of the
$X(3915)$ in a $D_s\bar{D}_s$ state is also calculated to be
slightly larger than the yield of the $X(3915)$ in a
$c\bar{c}s\bar{s}$ state, and the yield of the $X(3915)$ in a
$D_s^{tot}\bar{D}_s^{tot}$ state at kinetic freeze-out is about
three times larger than that of the $X(3915)$ in a $D_s\bar{D}_s$
state at chemical freeze-out.

The yield of the $X(3915)$ in a charmonium state is found to be
larger by about a factor of two compared to that in a thermal
model. Considering the constituent mass of charm quarks, 1500 MeV,
one can expect that the yield of the $X(3915)$ in a $c\bar{c}$
state is closer to that of the $J/\psi$ meson, thereby larger than
the yield of the $X(3915)$ in a thermal model due to its mass
3922.1 MeV. Even though the production of the $X(3915)$ is
suppressed owing to the internal relative momentum, the $2p$-wave,
the yield of the $X(3915)$ in a charmonium state is found to be
still larger than the expectation from the thermal model.

On the other hand, the yield of the $X(3915)$ in a compact
tetraquark state is evaluated to be smaller by about a factor of
forty compared to that in a thermal model, supporting the scenario
for the suppressed production of the multiquark hadrons by
coalescence in heavy ion collisions \cite{Cho:2011ew, Cho:2010db,
Cho:2017dcy}. Moreover, it is worthwhile to note that the yield of
the $X(3915)$ in a hadronic molecular state is also calculated to
be smaller than that in a thermal model, different from most
exotic hadrons except the $D_{s0}^*(2317)$, another charm-strange
meson \cite{Cho:2011ew, Cho:2010db, Cho:2017dcy}.

Therefore, together with the transverse momentum distributions of
the $X(3915)$ meson as shown in Fig. \ref{dNdpTX3915}, it is
expected that the yield of the $X(3915)$ meson for proposed states
in Table~\ref{X3915yields} would be helpful in specifying the
internal structure of the $X(3915)$ meson among its proposed
states from the measurement of its production in heavy ion
collisions.

\section{Conclusions}

We have investigated the $X(3915)$ as a $c\bar{c}s\bar{s}$ state in a quark model and studied its production in heavy ion collisions using a coalescence model.
Within the quark model, we analyze the internal color-spin structure of the $X(3915)$ together with its spatial wave function and find that the dominant component of the ground state corresponds to a well separated $D_s \bar{D}_s$ configuration.
This result suggests that the $X(3915)$ more likely corresponds to a hadronic molecular state, dominated by long range interactions beyond the description of the quark model.
Nevertheless, given the uncertainties of the quark model, a compact tetraquark configuration cannot be conclusively ruled out.

Along with the analysis on the quark model, we have also studied
the production of the $X(3915)$ meson, focusing on its production
by coalescence in relativistic heavy ion collisions at
$\sqrt{s_{NN}}=5.02$ TeV at LHC. Considering three proposed states
for the $X(3915)-$a charmonium, a tetraquark, and a hadronic
molecular state$-$we have evaluated the yields and transverse
momentum distributions of the $X(3915)$ for each state, and have
investigated the possibilities of its production in heavy ion
collisions. Moreover, we have considered two scenarios for the
$X(3915)$ in a hadronic molecular state, produced at both chemical
and kinetic freeze-outs.

The yield of the $X(3915)$ in a charmonium state is found to be
larger compared to those of the $X(3915)$ in both a hadronic
molecular and a tetraquark state, and even larger than the yield in
a thermal model by about a factor of two. On the other hand, the
yields in both a hadronic molecular state and a tetraquark state
are smaller compared to that in a thermal model, reflecting the
suppressed production of the multiquark hadrons by coalescence in
heavy ion collisions.

We find that the yield of the $X(3915)$ in a $D_s\bar{D}_s$ state
formed right after $D_s$ mesons are produced at chemical
freeze-out is slightly larger than the yield of the $X(3915)$ in a
$c\bar{c}s\bar{s}$ state, and the yield of the $X(3915)$ in a
$D_s^{tot}\bar{D}_s^{tot}$ state at kinetic freeze-out is larger
by about a factor of three compared to that of the $X(3915)$ in a
$D_s\bar{D}_s$ state at chemical freeze-out. It is noticeable to
see the smaller yields of the $X(3915)$ in both a hadronic
molecular and a tetraquark state compared to that in a thermal
model, similar to the case for the $D_{s0}^*(2317)$ meson.

Therefore, investigating the $X(3915)$ meson based on both the quark model and its
production in heavy ion collisions presents us with chances to understand the structure of
the $X(3915)$ in more detail. We expect that the observation of the
$X(3915)$ meson in relativistic heavy ion collisions
would help identify the internal structure of the $X(3915)$ meson
in the near future.

\section*{Acknowledgements}

This work was supported by the National Research Foundation of
Korea (NRF) grant funded by the Korea government (MSIT) No.
RS-2023-00280831, No. RS-2025-23963552, No.
2023R1A2C300302311 and No. 2023K2A9A1A0609492411.

\end{document}